\documentclass[useAMS,usegraphicx,usenatbib]{mn2e}

\usepackage{graphicx}
\usepackage{times}
\usepackage{amssymb}
\usepackage{amsmath}
\newif\ifAMStwofonts
\AMStwofontstrue

\def\rbs315{RBS\,315}
\def\pmna0525{PMN\,J0525$-$3343}
\def\rx1028{RX\,J1028.6$-$0844}
\def\gb1428{GB\,B1428$+$4217}
\def\pmnb1451{PMN\,J1451$-$1512}
\def\pksa1830{PKS\,B1830$-$211}
\def\pksb2126{PKS\,2126$-$0158}

\def\asca{{\emph{ASCA}}}
\def\bepposax{{\emph{BeppoSAX}}}

\def\rosat{{\emph{ROSAT}}}

\def\xmm{{\emph{XMM-Newton}}}

\def\cloudy{{\textsc{cloudy}}}
\def\arfgen{{\textsc{arfgen}}}
\def\rmfgen{{\textsc{rmfgen}}}
\def\sas{{\textsc{sas}}}
\def\wabs{{\textsc{wabs}}}
\def\xspec{{\textsc{xspec}}}
\def\zwabs{{\textsc{zwabs}}}
\def\nhi{\hbox{$N_{\mathrm{H}\thinspace\textsc{i}}$}}
\def\nh{\hbox{$N_{\mathrm{H}}$}}
\def\nhcold{\hbox{$N^{\mathrm{cold}}_{\mathrm{H}}$}}
\def\nhwarm{\hbox{$N^{\mathrm{warm}}_{\mathrm{H}}$}}
\def\nhgal{\hbox{$N^{\mathrm{Gal}}_{\mathrm{H}}$}}

\def\h{\hbox{$\mathrm{~h}$}}
\def\d{\hbox{$\mathrm{~d}$}}

\def\ghz{\hbox{$\mathrm{~GHz}$}}

\def\a{\hbox{$\mathrm{~\AA}$}}

\def\arcsec{\hbox{$\mathrm{~arcsec}$}}

\def\pcmsq{\hbox{$\mathrm{~cm^{-2}}$}}
\def\kmpspmpc{\hbox{$\mathrm{~km~s^{-1}~Mpc^{-1}}$}}

\def\kev{\hbox{$\mathrm{~keV}$}}

\def\mjy{\hbox{$\mathrm{~mJy}$}}

\def\ergpcmsqps{\hbox{$\mathrm{~erg~cm^{-2}~s^{-1}}$}}
\def\ergps{\hbox{$\mathrm{~erg~s^{-1}}$}}

\title{Radio and X-ray observations of an exceptional radio flare in the extreme z\,=\,4.72 blazar GB\,B1428+4217}
\author[M. A. Worsley et al.]
{\parbox[]{6.in}
{M.~A. Worsley,$^{1}$ 
A.~C. Fabian$^{1}$\thanks{E-mail: acf@ast.cam.ac.uk}, 
G.~G. Pooley$^{2}$ 
and C.~J. Chandler$^{3}$}\\
\footnotesize
$^{1}$Institute of Astronomy, Madingley Road, Cambridge CB3 0HA\\
$^{2}$Mullard Radio Astronomy Observatory, Cavendish Laboratory, Madingley Road, Cambridge CB3 0HE\\
$^{3}$National Radio Astronomy Observatory, 1003 Lopezville Road, PO Box 0, Socorro, New Mexico 87801, USA
}

\begin{document}
\maketitle

\label{firstpage}
\begin{abstract}
  We report on the extreme behaviour of the high redshift blazar \gb1428 at
  \hbox{$z=4.72$}. A continued programme of radio measurements has revealed an
  exceptional flare in the lightcurve, with the \hbox{$15.2\ghz$} flux density
  rising by a factor \hbox{$\sim3$} from \hbox{$\sim140\mjy$} to
  \hbox{$\sim430\mjy$} in a rest-frame timescale of only \hbox{$\sim4$} months
  -- much larger than any previous flares observed in this source. In addition
  to new measurements of the \hbox{$1.4$--$43\ghz$} radio spectrum we also
  present the analysis and results of a target-of-opportunity X-ray observation
  using \xmm, made close to the peak in radio flux. Although the X-ray data do
  not show a flare in the high energy lightcurve, we are able to confirm the
  X-ray spectral variability hinted at in previous observations.  \gb1428\ is
  one of several high-redshift radio-loud quasars that display a low energy
  break in the X-ray spectrum, probably due to the presence of excess absorption
  in the source. X-ray spectral analysis of the latest \xmm\ data is shown to be
  consistent with the warm absorption scenario which we have hypothesized
  previously. Warm absorption is also consistent with the observed X-ray
  spectral variability of the source, in which the spectral changes can be
  successfully accounted-for with a fixed column density of material in which
  the ionization state is correlated with hardness of the underlying power-law
  emission.
\end{abstract}


\section{Introduction}
Blazars are some of the most extreme extragalactic sources known: powerful,
radio-loud active galactic nuclei (AGN) in which the relativistic jet is aligned
with the observer's line of sight. Their spectral energy distribution (SED) is
dominated by beamed, non-thermal continuum emission from the jet and has a
characteristic double-humped shape, with synchrotron radiation from the jet
dominating at low energies, and Compton up-scattered emission dominating at high
energies. In BL Lac objects (the lower-luminosity class of blazars), the beamed
synchrotron emission can peak in the soft X-ray regime, with the second peak
dominating the TeV range. For flat-spectrum radio-quasars (FSRQs; the most
luminous blazars), the synchrotron radiation tends to peak in the infrared, with
the high energy emission peaking in the $\gamma$-ray regime \citep{fossati98}
and hard (\hbox{$\Gamma\sim1.5$}) power-law in the X-ray range.

In addition to their extreme luminosity and characteristic SED, blazars often
show spectacular broad-band variability, in both absolute brightness and
spectral shape. Flares in flux of several tens of per cent have been observed in
the radio to TeV regimes in several sources, on timescales of minutes to years
depending on wavelength. Blazars show strong, core-dominated and
highly-polarised radio emission, strongly supportive of the relativistic jet
interpretation.  Blazars are usually unresolved although very long baseline
interferometry (VLBI) can often reveal superluminal motion on milliarcsecond
scales within the nucleus, again consistent with the presence of a relativistic
jet.

\gb1428, at a redshift of \hbox{$z=4.72$}, shows an unexpected flattening of
the X-ray power-law continuum at soft X-ray energies. This was first seen in
\rosat\ and \bepposax\ observations \citep{boller00,fabian01b} and subsequently
confirmed with \xmm\ \citep{worsley04b}. A flattening in the X-ray spectral
index below observed energies of \hbox{$\sim1\kev$} has been observed in several
other high-redshift, radio-loud sources, e.g.: \rbs315\ at \hbox{$z=2.69$}
\citep{piconcelli05b}; \pmna0525\ at \hbox{$z=4.41$}
\citep{fabian01a,worsley04a}; \rx1028\ at \hbox{$z=4.28$}
\citep{yuan00,yuan05a}; \pmnb1451\ at \hbox{$z=4.76$} \citep{yuan05b};
\pksa1830\ at \hbox{$z=2.51$} \citep{derosa05}; and \pksb2126\ at
\hbox{$z=3.27$} \citep{fiore03,ferrero03}.  Although low energy miscalibrations
in \asca\ were initially proposed as an explanation \citep{grupe04}, the large
number of high-quality \xmm\ observations have shown beyond doubt that the
effect is real in several objects; that said, a number of \xmm\ observations
have also shown that several other high-redshift and radio-loud quasars show no
evidence of spectral flattening \citep{fiore03,grupe04,grupe06,piconcelli05a}. A
trend toward spectral flattening with increasing redshift has also been seen in
several studies of radio-loud objects
\citep[e.g.][]{cappi97,fiore98,reeves00,bassett04}.

The origin of the spectral depression is not fully understood, and, whilst the
spectral shape of \gb1428\ (and other sources) has been repeatedly characterised
using an absorbed power-law model, there is not yet sufficient evidence to be
certain that intrinsic absorption is the solution.  Both Galactic and
intergalactic absorption can be effectively discarded as viable solutions given
the high column densities required and, critically, by the fact that the
phenomenon has not been observed in radio-quiet objects at comparable redshifts
\citep{shemmer05,vignali05}, which would be expected to be similarly affected by
absorption systems lying along the line of sight
\citep{oflaherty97,fabian01a,worsley04a}. The remaining possibilities are
absorption intrinsic to the source or an underlying spectral break in the
power-law continuum; however, neither explanation can yet be ruled-out in any of
the cases of soft X-ray flattening.

As discussed by \citet{fabian01a}, an underlying break in the continuum could
arise in a number of situations. The hard X-ray component of the blazar SED is
believed to be due to inverse Compton scattering of seed photons by electrons in
the relativistic jet. A spectral break can arise through a low-energy cut-off in
the electron population or by a sharply peaked seed photon distribution. Both
mechanisms are rather speculative given our limited knowledge of the processes
which are important in blazar emission and difficult to reconcile with the
sharpness of the X-ray break, which occurs over a range of only a few keV in the
rest-frame. An intrinsic absorber, with a column density of
\hbox{$\nh\sim10^{22}$--$10^{23}\pcmsq$}, remains the most self-consistent
explanation for the soft X-ray flattening although very little is known about
the nature or origin of the material, which, given the lack of the effect in
radio-quiet sources, seems to be closely linked to the presence of jets in these
young AGN. X-ray data alone have been unsuccessful in breaking the degeneracy
between a cold (neutral) absorber or one that is warm (ionized). In some cases,
e.g. for \pmna0525\ and \rx1028, optical observations have been able to rule-out
cold absorption from the lack of a Lyman-limit system at the redshift of the
quasar by the presence of emission at rest-frame wavelengths shortward of
\hbox{$912\a$}. The lack of such a system in an object means the column density
in neutral hydrogen must be \hbox{$\nhi\lesssim3\times10^{17}\pcmsq$},
significantly less than the X-ray implied
\hbox{$\nhi\sim10^{22}$--$10^{23}\pcmsq$}.

In this paper we present the analysis and results from continued radio
monitoring of \gb1428, in particular, the details of an exceptional flare in the
lightcurve which has seen the \hbox{$15.2\ghz$} flux increase by a factor of
\hbox{$\sim3$} in only \hbox{$\sim4$} months in the rest-frame of the source. We
also report the findings of a recent \xmm\ target-of-opportunity observation
performed close to the peak in the radio lightcurve and discuss the X-ray
luminosity and spectral evolution displayed by this source. A broad-band radio
spectrum is also presented. We adopt \hbox{$H_{0}=71\kmpspmpc$},
\hbox{$\Omega_{\mathrm{M}}=0.27$} and \hbox{$\Omega_{\Lambda}=0.73$}
\citep{spergel03}. Errors are quoted at the 1$\sigma$ ($68$ per cent) level
unless stated otherwise. Also note that any X-ray fluxes quoted are always
corrected for Galactic absorption but are not corrected for any intrinsic
absorption unless stated otherwise.

\section{Previous observations}

\gb1428\ was discovered by \citet{hook98} as part of survey to identify FSRQs
with red optical counterparts (\gb1428\ has \hbox{$R=20.9$} with
\hbox{$B-R>3.40$}). It is at a redshift of \hbox{$z=4.72$}, making it one of the
most distant X-ray sources known. The blazar has an inverted radio-spectrum,
with a \hbox{$1.4$--$9\ghz$} spectral index of \hbox{$\alpha_{1.4-9}=-0.4$}
(\hbox{$f_{\nu}\propto\nu^{-\alpha}$}). Imaging revealed a core-dominated source
with a high level of polarisation (\hbox{$5.5$} per cent at \hbox{$1.5\ghz$};
\citealp{condon98}). The optical spectrum reveals broad emission lines due to
Ly$\alpha$, N\thinspace\textsc{v}, the
Si\thinspace\textsc{iv}/O\thinspace\textsc{iv}] blend and C\thinspace\textsc{iv}
\citep{hook98}. We will first summarise the findings of previous X-ray and radio
observations, before we move on to present the latest results and a joint
discussion of the spectrum and variability.

\subsection{X-ray spectrum}
\label{x-ray_spectrum}

An X-ray counterpart was found in an archival \rosat\ Position Sensitive
Proportional Counter (PSPC) observation taken in 1992. Subsequent observations
were performed with the \rosat\ High Resolution Imager (HRI; \citealp{fabian97})
and \asca\ \citep{fabian98}. The inferred isotropic X-ray luminosity exceeded
\hbox{$10^{47}\ergps$} which suggested highly beamed emission. The
\hbox{$0.5$--$10\kev$} data were consistent with a hard power-law spectrum, with
a photon index of \hbox{$\Gamma=1.29\pm0.05$}, and an upper limit to any
intrinsic absorption at the redshift of the quasar of
\hbox{$4.5\times10^{22}\pcmsq$} ($90$ per cent confidence); however, the authors
noted that the SIS instrument alone did point to some excess absorption -- a
column density of \hbox{$(1.3\pm0.6)\times10^{23}\pcmsq$} for
\hbox{$\Gamma=1.43\pm0.12$}. Excess absorption was confirmed with a further
\rosat\ PSPC observation, which found a power-law index of
\hbox{$\Gamma=1.4\pm0.2$} and a column density of
\hbox{$\nhcold=(1.52\pm0.28)\times10^{22}\pcmsq$} ($90$ per cent errors) at
\hbox{$z=4.72$} \citep{boller00}. Further observations with \bepposax\ also
measured \hbox{$\Gamma=1.45\pm0.10$} power-law and excess absorption of
\hbox{$\nhcold=7.8^{+8.7}_{-6.0}\times10^{22}\pcmsq$} ($90$ per cent errors;
\citealp{fabian01b}).

Warm absorption was first suggested by \citet{fabian01b} following the
\bepposax\ observations. In addition to modelling the results of that
observation, the authors also fitted the previous \asca\ and \rosat\ data. A
marginally consistent set of parameters were found with a column density of
\hbox{$(2$--$3)\times10^{23}\pcmsq$} and an ionization parameter
\hbox{$\xi\sim300$}.\footnote{\citet{fabian01b} quote
  \hbox{$\xi_{1-1000~\mathrm{Ryd}}$}, defined as
  \hbox{$L_{1-1000~Ryd}/n_{\mathrm{H}}R^{2}$}, where
  \hbox{$L_{1-1000~\mathrm{Ryd}}$} is the \hbox{$1$--$1000$} Rydberg luminosity,
  \hbox{$n_{\rm{H}}$} is the space density of H\thinspace\textsc{i} and $R$ is
  the distance from the source. For \gb1428, the \citet{fabian01b} definition is
  a factor \hbox{$\sim2$} higher than the
  \hbox{$\xi=L_{2-10~\mathrm{keV}}/n_{\mathrm{H}}R^{2}$} definition used in this
  paper.}  \xmm\ observations \citep{worsley04b} confirmed the presence of soft
X-ray spectral flattening, with a column density of
\hbox{$\nhcold\sim(1.4$--$1.6)\times10^{22}\pcmsq$} if cold, or up to
\hbox{$\nhwarm\sim10^{23}\pcmsq$} with an ionization parameter
\hbox{$\xi\sim100$}, if due to a warm absorber. Interestingly, the \xmm\ data
showed a distinct change in the spectral index of the source, both from the
earlier observations as well in the \hbox{$\sim7\d$} (rest-frame) between the
two \xmm\ exposures. The first observation indicated \hbox{$\Gamma\sim1.9$},
whilst the second found \hbox{$\Gamma\sim1.75$}. Whilst the source had long been
known to show extreme luminosity variations, this was the first definitive
example of spectral changes.

\subsection{X-ray variability}
\label{x-ray_variability}

No clear variability was noted between the original \rosat\ PSPC detection in
1992 and the \rosat\ HRI and \asca\ observations during 1996 and 1997: all were
consistent with a \hbox{$0.1$--$2.4\kev$} soft X-ray flux\footnote{To reiterate,
  all X-ray fluxes have been corrected for Galactic absorption only, and not for
  any apparently intrinsic absorption.} of \hbox{$\sim10^{-12}\ergpcmsqps$},
with no significant evidence for any differences \citep{fabian97,fabian98}.
However, a series of four \rosat\ HRI observations from 1997 December 12 until
1998 January 23 showed substantial X-ray variability, with the
\hbox{$0.1$--$2.4\kev$} flux increasing from \hbox{$(0.6\pm0.1)\times10^{-12}$}
to \hbox{$(1.1\pm0.1)\times10^{-12}\ergpcmsqps$} between the third and fourth
exposure -- a period of only \hbox{$2.4\d$} in the rest-frame of the source
\citep{fabian99}. Subsequent \rosat\ PSPC observations then identified
\hbox{$25$} per cent variability on a rest-frame timescale of \hbox{$\sim1.8\h$}
\citep{boller00}. Furthermore, the mean \hbox{$0.1$--$2.4\kev$} flux during the
observation was \hbox{$(2.7\pm0.6)\times10^{-12}\ergpcmsqps$}, 3 times the
previous \rosat\ and \asca\ measurements in this band. A cautionary note must be
attached to these PSPC measurements, which were amongst the final observations
of the \rosat\ satellite which experienced some detector anomalies and the
development of a large gain hole in the PSPC instrument, although careful data
reduction techniques \citep[refer to][]{boller00} should mean the results were
not significantly affected. The \bepposax-measured \citep{fabian01a}
\hbox{$2$--$10\kev$} flux and spectral shape imply a \hbox{$0.1$--$2.4\kev$}
flux of \hbox{$\sim10^{-12}\ergpcmsqps$}, indicating a decrease back to the
level seen prior to the \rosat\ PSPC observation. The two \xmm\ exposures of
2002 December 09 and 2003 January 17 measured \hbox{$0.1$--$2\kev$} fluxes of
\hbox{$(1.17\pm0.04)\times10^{-12}$} and
\hbox{$(0.85\pm0.02)\times10^{-12}\ergpcmsqps$} respectively, consistent with
other measurements but revealing a \hbox{$\sim30$} per cent decrease in the
\hbox{$7\d$} (rest-frame) between the \xmm\ observations.

In the hard X-ray band, \asca\ and \bepposax\ \citep{fabian98,fabian01a}
measured consistent \hbox{$2$--$10\kev$} fluxes of \hbox{$2.5\times10^{-12}$}
and \hbox{$2.76\times10^{-12}\ergpcmsqps$}. The two \xmm\ exposures measured
\hbox{$2$--$10\kev$} fluxes of \hbox{$(1.21\pm0.06)\times10^{-12}$} and
\hbox{$(1.08\pm0.03)\times10^{-12}\ergpcmsqps$} respectively. These were
\hbox{$\sim2$--$2.5$} times lower than the earlier measurements and indicating a
\hbox{$10$} per cent decrease in only \hbox{$7\d$} (rest-frame) between the
\xmm\ observations (note that this decrease is much less than the
\hbox{$\sim30$} per cent fall seen for the soft X-ray band because of the
distinct change in the spectral shape of the source between the two
observations: see Section~\ref{x-ray_spectrum}). We will present soft and hard
band X-ray lightcurves shortly, following the details and results of the new
\xmm\ measurements.

\subsection{Radio properties}

\citet{fabian97} collated various radio measurements for \gb1428\ and present
the SED of the source, noting the flat radio spectrum, high level of
polarisation and indications of \hbox{$5\ghz$} variability. An intensive
monitoring programme was carried out between 1998 April and December using the
Ryle Telescope at Cambridge at \hbox{$15.2\ghz$}. The programme formed part of a
radio and X-ray variability study, although the radio observations were somewhat
later than the \rosat\ HRI exposures carried out in 1997 December and 1998
January \citep{fabian99}. The Ryle data revealed significant \hbox{$15.2\ghz$}
variability, with a flux increase from \hbox{$\sim100\mjy$} to
\hbox{$\sim140\mjy$} in a rest-frame period of only \hbox{$\sim20\d$}. The flare
ended abruptly mid-way through the observing programme, with the flux decreasing
back to \hbox{$\sim105\mjy$} within a fraction of a day in the rest-frame. The
lightcurve also shows strong variability at the \hbox{$15$} per cent level with
a rest-frame timescale of only \hbox{$\sim2\d$}.

\section{New radio observations}

\begin{figure}
\rotatebox{270}{
\resizebox{!}{\columnwidth}
{\includegraphics{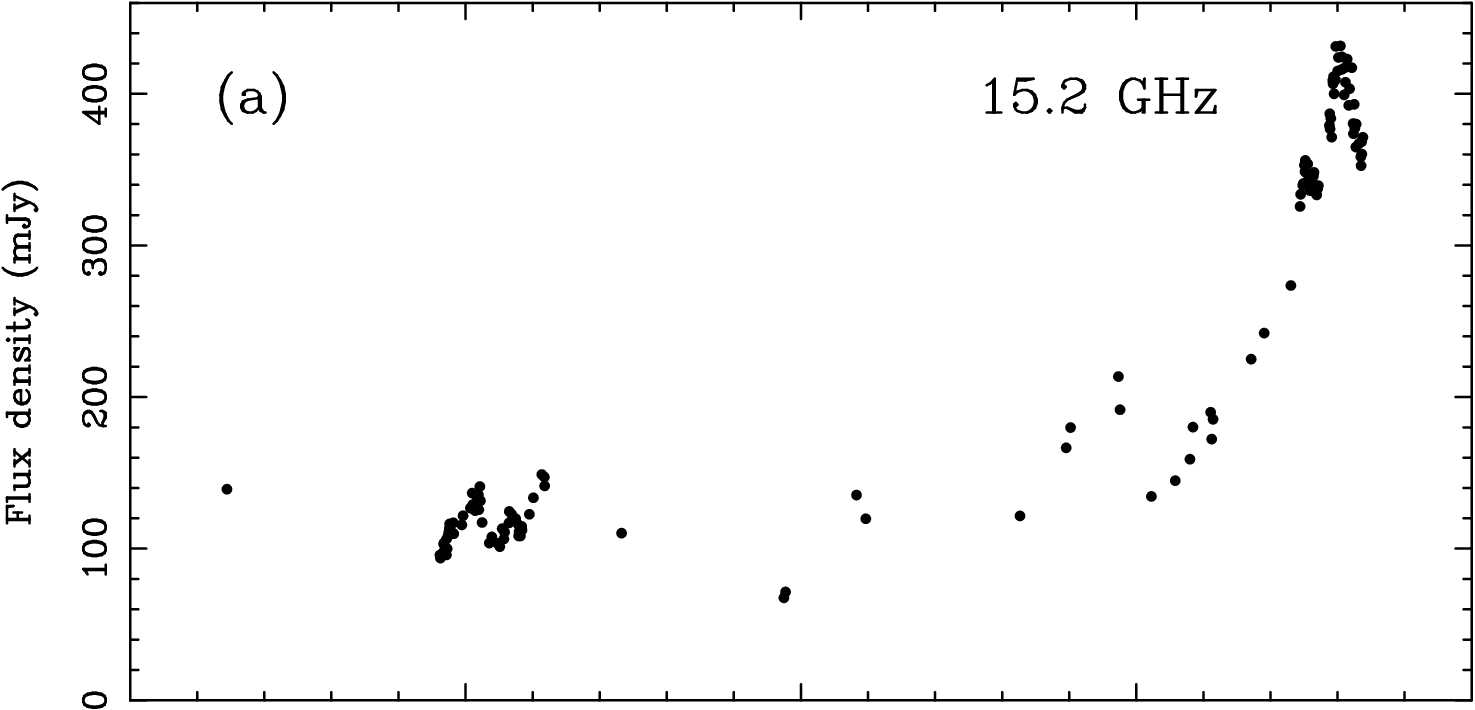}}}
\rotatebox{270}{
\resizebox{!}{\columnwidth}
{\includegraphics{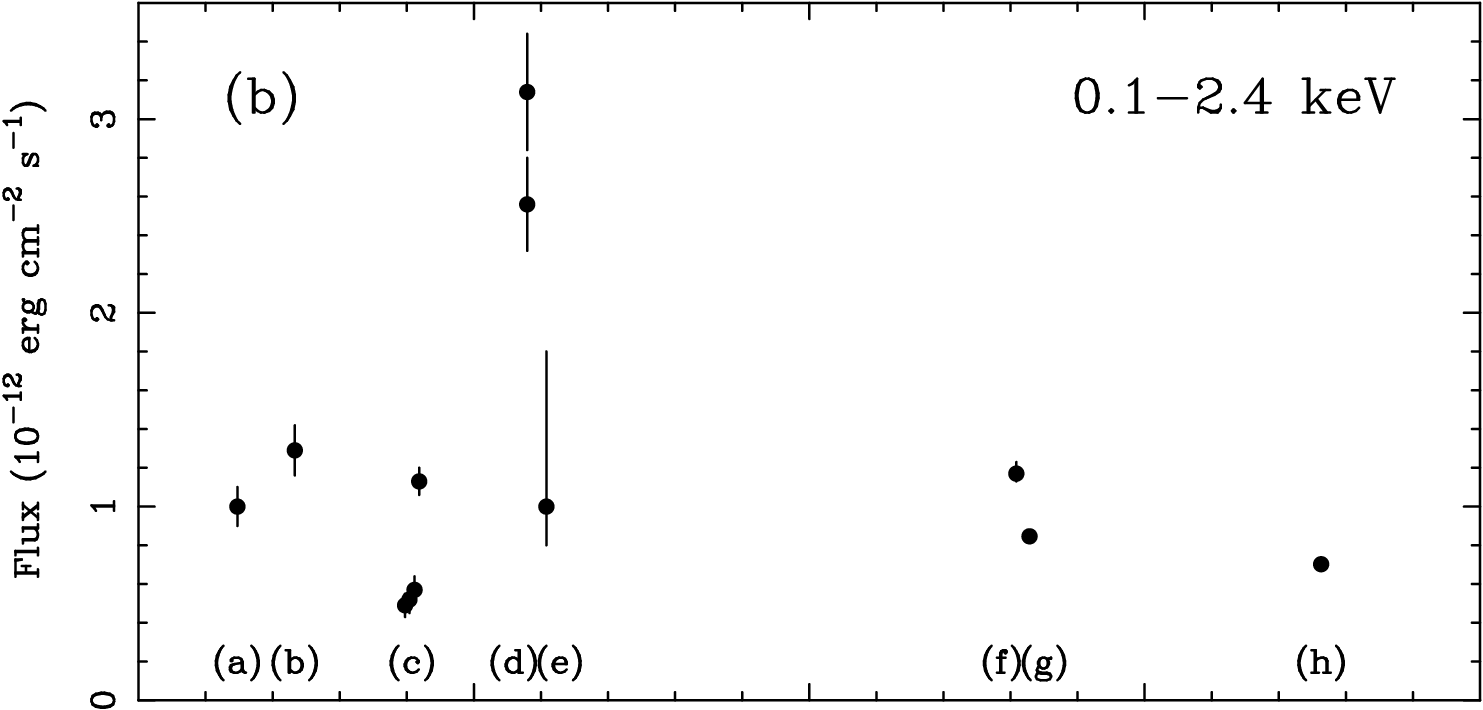}}}
\rotatebox{270}{
\resizebox{!}{\columnwidth}
{\includegraphics{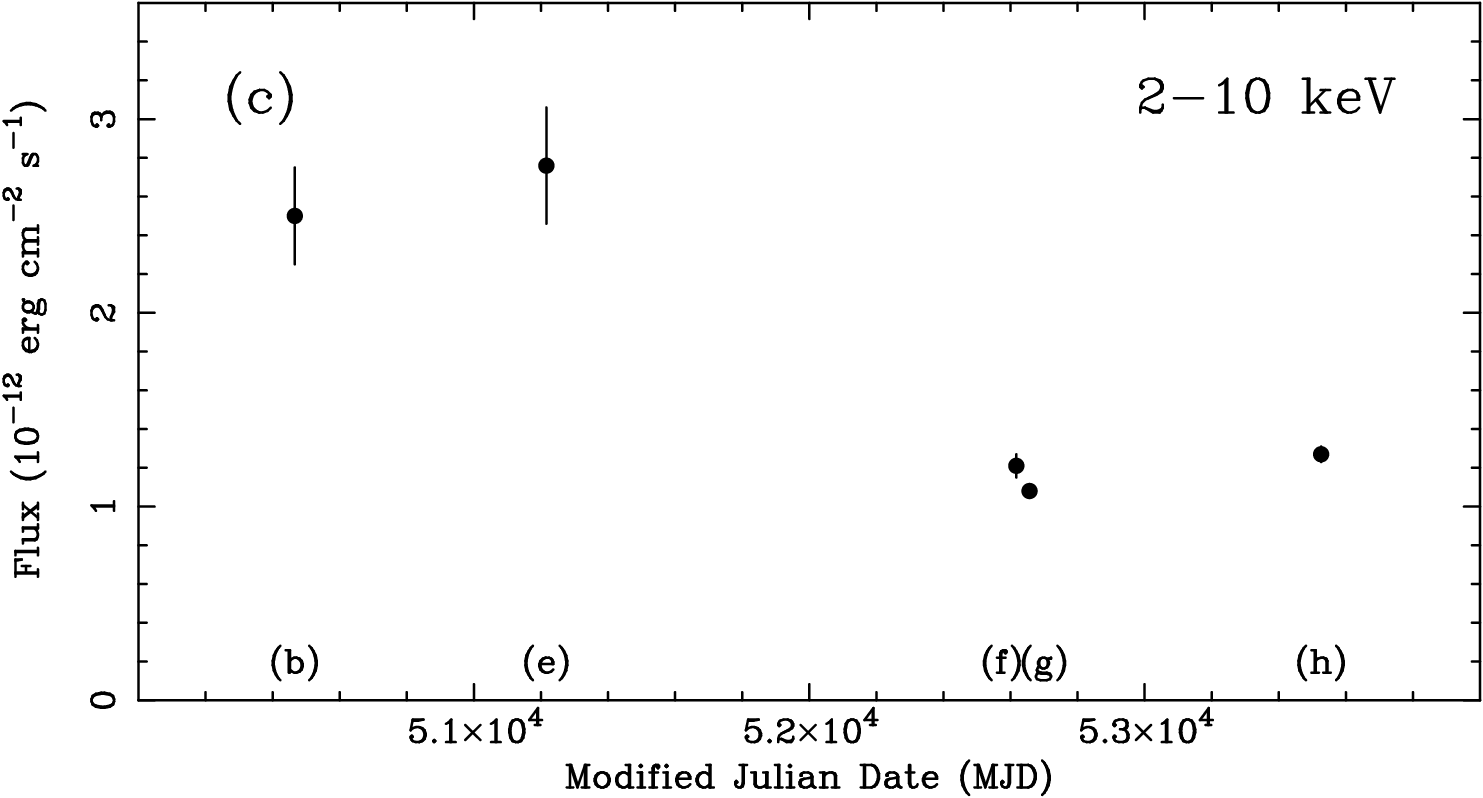}}}
\caption{(a) The \hbox{$15.2\ghz$} (rest-frame \hbox{$86\ghz$} radio lightcurve
  of \gb1428\ as measured by the Ryle Telescope in Cambridge from 1998 March 20
  (MJD 50923) until 2005 November 01 (MJD 53623). (b) The soft
  \hbox{$0.1$--$2.4\kev$} (rest-frame \hbox{$\sim0.6$--$14\kev$}) X-ray
  lightcurve for the same time period. (c) The hard \hbox{$2$--$10\kev$}
  (rest-frame \hbox{$\sim11$--$57\kev$}) X-ray lightcurve.  X-ray Measurements
  and errors have been taken directly from the reference paper or, where
  possible, estimated from the spectral parameters and other details. References
  (in order of observation date): (a) \rosat\ HRI \citep{fabian97}; (b) \asca\
  \citep{fabian98}; (c) \rosat\ HRI \citep{fabian99}; (d) \rosat\ PSPC
  \citep{boller00}; (e) \bepposax\ \citep{fabian01b}; (f) \xmm\ rev. 549
  \citep{worsley04b}; (g) \xmm\ rev. 569 \citep{worsley04b}; (h) \xmm\ rev. 1005
  (this paper).}
\label{lightcurves}
\end{figure}

The Ryle telescope has been used to measure the \hbox{$15.2\ghz$} radio flux of
\gb1428\ on a regular basis since the monitoring programme in 1997-8 \citep[for
details of the observing method see][]{pooley97}. The lightcurve is shown in
Fig.~\ref{lightcurves}(a). Measurements after the 1997-8 programme, although
sparse, indicated no significant behaviour beyond the variability that had
already been observed until that point. The flux decreased to only
\hbox{$70\mjy$} before starting to climb. In less than 6 months (rest-frame),
the flux peaked at \hbox{$\sim210\mjy$} before dropping suddenly to
\hbox{$140\mjy$}. The source then began to steadily grow in flux -- the start of
a flare of a enormous magnitude that has continued to the present. By 2005
August, the \hbox{$15.2\ghz$} flux had reached \hbox{$\sim430\mjy$}, a factor
\hbox{$\sim3$} increase in only \hbox{$\sim4$} months (rest-frame). The
lightcurve continues to show intra-day variations of \hbox{$\sim10$} per cent,
with several `micro-flares' visible in the primary flare.

\begin{figure}
\rotatebox{270}{
\resizebox{!}{\columnwidth}
{\includegraphics{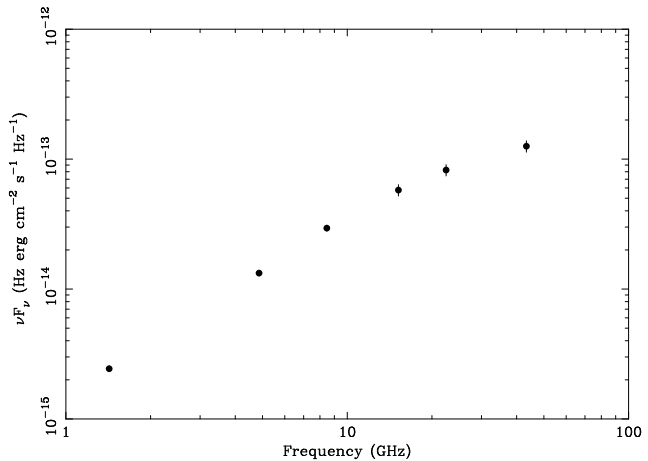}}}
\caption{A \hbox{$\nu F_{\nu}$} plot showing the radio spectrum of the blazar
  using data from the VLA and Ryle Telescope. The data are given in
  Table~\ref{radio_sed_data}.}
\label{radio_sed}
\end{figure}

A radio spectrum was obtained on 2005 October 14 with the Very Large Array (VLA)
at \hbox{$1.425$}, \hbox{$4.86$}, \hbox{$8.46$}, \hbox{$22.46$} and
\hbox{$43.34\ghz$}, and almost simultaneously at \hbox{$15.2\ghz$} with the Ryle
Telescope, shortly after the peak flux density at \hbox{$15.2\ghz$} (flux
densities at this frequency are of Stokes' $I+Q$; all VLA measurements are of
Stokes' $I$). The data and spectrum are given in Table~\ref{radio_sed_data} and
Fig.~\ref{radio_sed}, respectively. The absolute flux calibration for the VLA
data was done using 3C286 and the measured flux densities at each frequency are
given in Table~\ref{radio_sed_data}. The flux measurements include an
uncertainty in the calibration of $5$ per cent for the three lowest frequencies
and $10$ per cent for the remainder.

\begin{table}
\centering
\caption{VLA measurements of the \gb1428\ radio spectrum from \hbox{$1.425$} to 
  \hbox{$43.4\ghz$} as measured on 2005 October 14 (MJD 53657), close to 
  the peak in the radio lightcurve. The \hbox{$15.2\ghz$} data are from 
  the Ryle Telescope on 2005 October 12. Ryle Telescope flux densities are of
  Stokes' $I+Q$ but the VLA measurements are of Stokes' $I$. The flux densities 
  of the calibration source 3C286 are also given.}
\label{radio_sed_data}
\begin{tabular}{lcr}
\hline
Frequency & \gb1428\           & 3C286 calibrator\\
(GHz)     & flux density (mJy) & flux density (Jy)\\
\hline
$1.425$   & $171\pm9$    & $14.70$ \\
$4.86$    & $273\pm14$   & $7.47$ \\
$8.46$    & $348\pm17$   & $5.20$ \\
$15.2$    & $380\pm38$   & $3.50$ \\
$22.46$   & $367\pm37$   & $2.52$ \\
$43.34$   & $290\pm29$   & $1.45$ \\
\hline
\end{tabular}
\end{table}

\section{New {\emph{XMM-Newton}} observations}

\subsection{Data reduction}

Following the extraordinary increase in \hbox{$15.2\ghz$} flux, \xmm\ was used to
perform a target-of-opportunity (ToO) X-ray observation co-incident with the
radio brightening. Various details of the observation, which was carried out
during \xmm\ revolution 1005, are given in Table~\ref{xmm_exposures}. The
European Photon Imaging Camera (EPIC) pn, MOS-1 and MOS-2 instruments were
active. Data were reduced using the \xmm\ Science Analysis System (\sas) version
6.4.0-6.5.0. Periods of background flaring were excluded using a lightcurve
extracted from the whole pn/MOS chip in each case. Raw events were filtered in
the recommended\footnote{{\tt{(FLAG==0)\&\&(PATTERN<=4)\&\&(PI IN [150,15000])}}
  and {\tt{(\&XMMEA\_EM)\&\&(PATTERN<=12)\&\&(PI IN [200,10000])}} for the pn
  and MOS cameras, respectively. See the \xmm\ Science Operations Centre
  document XMM-SOC-CAL-TN-0018.} way.

\begin{table*}
\centering
\caption{Details of the ToO \xmm\ observation from revolution 1005 along with
the details of the previous exposures of revolutions 549 and 569.}
\label{xmm_exposures}
\begin{tabular}{lcccccccccr}
\hline
Revolution & Date & MJD & Mode & Filter & \multicolumn{3}{c}{Good Exposure Times (ks)} & \multicolumn{3}{c}{Good counts} \\
&&&&& pn & MOS-1 & MOS-2 \\
\hline
549  & 2002 Dec 09 & 52617.3 & Full frame & Thin & 2.6  & 4.1  & 4.3  & 2162 & 996 & 979 \\
569  & 2003 Jan 17 & 52656.7 & Full frame & Thin & 11.5 & 14.2 & 14.2 & 6797 & 2577 & 2541 \\
1005 & 2005 Jun 05 & 53526.3 & Full frame & Thin & 12.9 & 15.5 & 15.9 & 7222 & 2604 & 2657 \\
\hline
\end{tabular}
\end{table*}

Spectra were produced from events extracted from circular regions centred on the
source. Radii of approximately \hbox{$45$} and \hbox{$75\arcsec$} were used
for the pn and MOS, respectively. Background spectra were formed using events
from nearby source-free regions on the same chip as the source. The regions used
had respective radii of approximately \hbox{$130$} and \hbox{$170\arcsec$} for
the pn and MOS. The \sas\ tasks \arfgen\ and \rmfgen\ were used to generate
the appropriate response matrices. Background-corrected spectra were produced
for each instrument and grouped to a minimum of \hbox{$20$} counts in each
energy channel.

\subsection{X-ray lightcurve}

Figs.~\ref{lightcurves}(b) and \ref{lightcurves}(c) show X-ray lightcurves to
match the radio data shown in Fig.~\ref{lightcurves}(a). Measurements of the
X-ray flux in the soft \hbox{$0.1$--$2.4\kev$} and hard \hbox{$2$--$10\kev$}
bands (corresponding to rest-frame energies of \hbox{$\sim0.6$--$14$} and
\hbox{$\sim11$--$57\kev$}, respectively) are shown.  The measurements are taken
from the various observations described in Section~\ref{x-ray_variability} and
the new \xmm\ data (2005 June 05). This observation had \hbox{$0.1$--$2.4$} and
\hbox{$2$--$10\kev$} fluxes of \hbox{$(0.70\pm0.02)\times10^{-12}$} and
\hbox{$(1.27\pm0.04)\times10^{-12}\ergpcmsqps$}, respectively. The X-ray
lightcurve illustrates the strong variations in flux that have been seen
throughout the various observations of the source (see
Section~\ref{x-ray_variability} for details and references). The latest \xmm\
data do not show any evidence of variability within the observation; however,
the \hbox{$0.1$--$2.4$} and \hbox{$2$--$10\kev$} fluxes are different to those
measured in the previous \xmm\ observations.

There appears to be no obvious correlation between the X-ray and radio
lightcurves, although the sparse sampling of the lightcurves does not allow us
to rule anything out. Interestingly, one would not necessarily expect a
correlation with zero time-lag, particularly since blazar emission is dominated
by beamed radiation from a relativistic jet. Continued monitoring in both radio
and X-ray wavebands over the coming years may reveal correlations which are not
clear with the present data.

\subsection{X-ray spectrum}

\xspec\ version 11.3 was used for spectral fitting, which was done for the pn,
MOS-1 and MOS-2 data jointly. A simple power-law model \footnote{Galactic cold
  absorption of \hbox{$\nhgal=1.40\times10^{20}\pcmsq$} \citep{elvis94}, applied
  using \wabs\ in \xspec\ \citep{morrison83}, is included in all spectral
  models}, fitted at energies \hbox{$>1\kev$}, is a good match to the spectrum,
with a reduced chi-squared of \hbox{$\chi^{2}_{\nu}=1.024$} for \hbox{$\nu=316$}
degrees of freedom (the probability of the model not representing the data,
\hbox{$1-\rm{Pr(H_{0})}$}, is \hbox{$\sim63$} per cent and there is therefore no
significant evidence to suggest the model is unacceptable).  However, when the
same model is applied to the full energy range, the fit is unacceptable (at the
\hbox{$99.998$} per cent level), with \hbox{$\chi^{2}_{485}=1.359$}.  Re-fitting
the simple power-law model to the full energy range also leads to an
unacceptable fit (at the \hbox{$99.7$} per cent level), with
\hbox{$\chi^{2}_{483}=1.187$}.

Soft X-ray spectral flattening is once again evident, and we use a power-law
plus intrinsic cold absorption model (using \zwabs\ in \xspec;
\citealp{morrison83}). A joint analysis was performed with the earlier \xmm\
data taken in 2002 December and 2003 January (\xmm\ revolutions 549 and 569),
respectively; the details are included in Table~\ref{xmm_exposures} \citep[for
further information on these data refer to][]{worsley04b}. Table~\ref{xmm_cold}
shows the various results.  Allowing the power-law photon index $\Gamma$, the
\zwabs\ column density \nhcold, and the normalization to vary between
observations can yield good fits to the revolution 549 and 569 data, although
the 1005 data is less satisfactory.  Fixing $\Gamma$ and \nhcold\ between the
observations results in an unacceptable fit (at the \hbox{$99.994$} per cent
level), with \hbox{$\chi^{2}_{1121}=1.170$}. Alternatively, allowing the
\nhcold\ to vary whilst $\Gamma$ is fixed is also unacceptable, with
\hbox{$\chi^{2}_{1119}=1.111$} (at the \hbox{$99.5$} per cent level): there is
clear spectral variability in the source (a discussion of which we will come to
shortly). Interestingly, fixing \nhwarm\ but allowing $\Gamma$ to vary leads to
a statistically acceptable fit, with \hbox{$\chi^{2}_{1119}=0.999$}.  All three
data-sets are explicable with variations in the underlying power-law slope, with
a fixed column density of \hbox{$(1.4\pm0.1)\times10^{22}\pcmsq$} of intrinsic
cold absorption. This model is not a good fit to the revolution 1005 data on its
own (\hbox{$\chi^{2}_{484}=1.095$}; unacceptable at the \hbox{$99.7$} per cent
level), something which may suggest that the spectrum is more complex, perhaps
involving warm rather than cold absorption.


\begin{table*}
\centering
\caption{Spectral fits to a power-law plus intrinsic cold absorption model
  (inclusive of Galactic cold absorption of
  \hbox{$\nhgal=1.40\times10^{20}\pcmsq$}). The fits use different combinations of
  parameters which were fixed to be the same for the separate observations, and
  those which were free to vary independently. The $\chi^{2}_{\nu}$ and degrees of 
  freedom $\nu$ is shown for the total fit and also when the model is applied to the 
  individual data set alone.}
\label{xmm_cold}
\begin{tabular}{lcccccr}
\hline
Revolution & $\Gamma$ & \nhcold\          & \multicolumn{2}{c}{Total} & \multicolumn{2}{c}{Individual} \\
           &          & ($10^{22}\pcmsq$) & $\chi^{2}_{\nu}$ & $\nu$ & $\chi^{2}_{\nu}$ & $\nu$ \\
\hline
\multicolumn{7}{l}{(a) $\Gamma$ and \nhcold\ independent:}\\
549  & $1.89\pm0.04$ & $1.61\pm0.28$ &         &        & $0.831$ & $185$ \\
569  & $1.76\pm0.02$ & $1.67\pm0.16$ & $0.991$ & $1117$ & $0.946$ & $454$ \\
1005 & $1.49\pm0.02$ & $0.95\pm0.15$ &         &        & $1.081$ & $484$ \\
\hline
\multicolumn{7}{l}{(b) $\Gamma$ and \nhcold\ fixed:}\\
549  &               &               &         &        & $1.189$ & $185$ \\
569  & $1.67\pm0.01$ & $1.40\pm0.10$ & $1.170$ & $1121$ & $1.005$ & $454$ \\
1005 &               &               &         &        & $1.308$ & $484$ \\
\hline
\multicolumn{7}{l}{(c) $\Gamma$ fixed; \nhcold\ independent:}\\
549  &               & $0.51\pm0.17$ &         &        & $1.031$ & $185$ \\
569  & $1.67\pm0.01$ & $1.15\pm0.13$ & $1.111$ & $1119$ & $0.989$ & $454$ \\
1005 &               & $2.12\pm0.15$ &         &        & $1.245$ & $484$ \\
\hline
\multicolumn{7}{l}{(d) $\Gamma$ independent; \nhcold\ fixed:}\\
549  & $1.86\pm0.27$ &               &         &        & $0.835$ & $185$ \\
569  & $1.73\pm0.02$ & $1.4\pm0.1$   & $0.999$ & $1119$ & $0.953$ & $454$ \\
1005 & $1.53\pm0.02$ &               &         &        & $1.095$ & $484$ \\
\hline
\end{tabular}
\end{table*}

Warm absorption is also a very plausible candidate for the spectral shape of the
source. Table~\ref{xmm_warm} shows the results of various power-law plus warm
absorption model fits to the data. The photoionization code \cloudy\
\citep{ferland98} was used to generate the warm absorber models. The ionization
parameter $\xi$ is defined from \hbox{$\xi=L/n_{\rm{H}}R^{2}$}; where $L$ is the
rest-frame \hbox{$2$--$10\kev$} luminosity, $n_{\rm{H}}$ is the absorber space
density and $R$ is the distance from the source.  As with the cold absorption
models we tested various combinations of free and fixed parameters in the
analysis. Allowing $\Gamma$ to vary along with both \nhcold\ and the ionization
parameter $\xi$ (this is a fixed \hbox{$\xi=0$} in the cold absorber models)
results in a good fit with \hbox{$\chi^{2}_{1114}=0.985$}. Fixing $\Gamma$
results in a poorer fit with \hbox{$\chi^{2}_{1127}=1.059$}, although the
evidence for an unacceptable fit is only at the \hbox{$91.7$} per cent level.
Allowing $\Gamma$ and $\xi$ to vary whilst maintaining a fixed \nhwarm\ provides
a very good description of the data, with \hbox{$\chi^{2}_{1126}=0.987$} and
reasonable values for the fits to the individual data-sets. A fixed column
density of absorption with ionisation state and underlying power-law variations
seems the more physically plausible situation and is also consistent with the
X-ray spectral variation which we will describe next.

\begin{table*}
\centering
\caption{Spectral fits to a power-law plus intrinsic warm absorption model
  (inclusive of Galactic cold absorption of
  \hbox{$\nhgal=1.40\times10^{20}\pcmsq$}). The fits use different combinations of
  parameters which were fixed to be the same for the separate observations, and
  those which were free to vary independently. The $\chi^{2}_{\nu}$ and degrees of 
  freedom $\nu$ is shown for the total fit and also when the model is applied to the 
  individual data set alone.}
\label{xmm_warm}
\begin{tabular}{lccccccr}
\hline
Revolution & $\Gamma$ & $\log_{10}(\nhwarm/\rm{cm^{-2}})$ & $\log_{10}(\xi/\rm{erg~cm~s^{-1}})$ & \multicolumn{2}{c}{Total} & \multicolumn{2}{c}{Individual} \\
           &          &                    &                & $\chi^{2}_{\nu}$ & $\nu$ & $\chi^{2}_{\nu}$ & $\nu$ \\
\hline
\multicolumn{8}{l}{(a) $\Gamma$, \nhwarm\ and $\xi$ independent:}\\
549  & $1.89\pm0.06$ & $22.34\pm0.21$ & $<-1$          &         &        & $0.834$ & $185$ \\
569  & $1.77\pm0.04$ & $22.37\pm0.15$ & $-0.04\pm1.04$ & $0.985$ & $1114$ & $0.945$ & $454$ \\
1005 & $1.55\pm0.04$ & $22.90\pm0.67$ & $2.03\pm0.44$  &         &        & $1.061$ & $484$ \\
\hline
\multicolumn{8}{l}{(b) $\Gamma$ fixed; \nhwarm\ and $\xi$ independent:}\\
549  &               & $21.82\pm0.50$ & $<-1$          &         &        & $0.992$ & $185$ \\
569  & $1.70\pm0.03$ & $22.24\pm0.13$ & $-0.04\pm0.96$ & $1.059$ & $1127$ & $0.969$ & $454$ \\
1005 &               & $>23$          & $1.84\pm0.26$  &         &        & $1.153$ & $484$ \\
\hline
\multicolumn{8}{l}{(c) $\Gamma$ and $\xi$ independent; \nhwarm\ fixed:}\\
549  & $1.91\pm0.06$ &                & $-0.96\pm0.30$ &         &        & $0.836$ & $185$ \\
569  & $1.77\pm0.02$ & $22.34\pm0.04$ & $-0.18\pm1.04$ & $0.987$ & $1126$ & $0.946$ & $454$ \\
1005 & $1.51\pm0.02$ &                & $1.44\pm0.15$  &         &        & $1.071$ & $484$ \\
\hline
\end{tabular}
\end{table*}

\section{X-ray spectral variability}

\begin{figure}
\rotatebox{270}{
\resizebox{!}{\columnwidth}
{\includegraphics{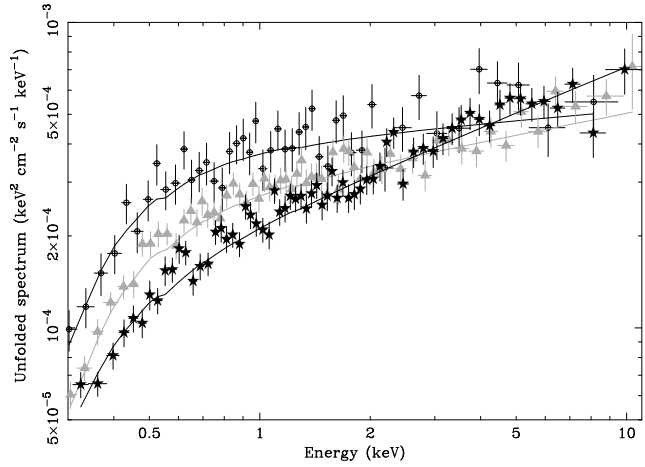}}}
\caption{Unfolded X-ray spectra, in \hbox{$\nu F_{\nu}$} space, from the \xmm\
    revolution 549 (open black circles), 569 (solid grey diamonds) and 1005
    (solid black stars) observations.  A power-law plus intrinsic cold
    absorption model was used to unfold the spectra and is shown as a solid line
    in each case.}
\label{eeufs}
\end{figure}

The X-ray lightcurves (Figs.~\ref{lightcurves}(a) and \ref{lightcurves}(b)) show
strong amplitude variations but also show differences between the
\hbox{$0.1$--$2.4$} and \hbox{$2$--$10\kev$} behaviour. The
\hbox{$(0.1$--$2.4)/(2$--$10)\kev$} flux ratio was \hbox{$\sim0.5$} and then
\hbox{$\sim0.4$} during the \asca\ and \bepposax\ observations. The ratio was
much larger during the later \xmm\ observations, with more evidence of spectral
variation between the revolution 549 and 569 observations, with the ratio
falling from \hbox{$0.97\pm0.06$} to \hbox{$0.78\pm0.03$}. The revolution 1005
data indicate a value of \hbox{$0.55\pm0.02$}, close to that seen in the earlier
observations. Here, we have used a simple soft/hard band flux ratio to
characterise the spectral variation but the evolution can also be well-described
by a change in the underlying power-law slope with a fixed column density of
absorption.  Fig.~\ref{eeufs} shows the X-ray spectra of the three \xmm\
observations and clearly indicates a hardening of the power-law photon index in
moving from the revolution 549 to the 569 and 1005 data.

\begin{figure}
\rotatebox{270}{
\resizebox{!}{\columnwidth}
{\includegraphics{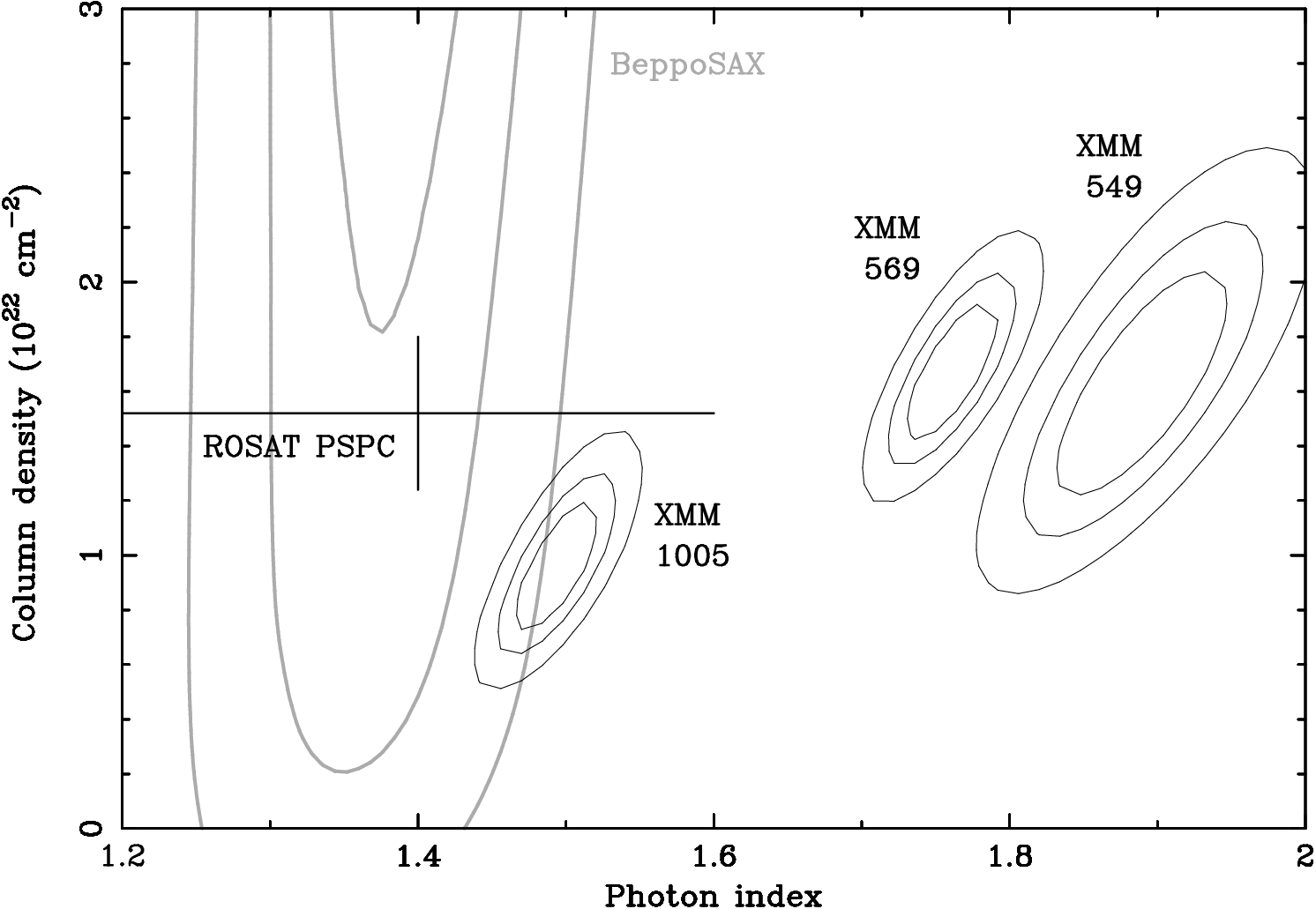}}}
\caption{Confidence contours ($68$, $90$ and $99$ per cent) in the photon index
  $\Gamma$ and the cold absorber column density \nhcold. The \bepposax\ and
  \rosat\ PSPC data \citep{fabian01b,boller00} are shown, along with results
  from \xmm\ revolutions 549, 569 and 1005 (\citealp{worsley04b}; this paper).
  The data are all marginally consistent with a constant \nhcold\ model in which
  $\Gamma$ varies widely, although the column densities implied by the
  revolution 569 and 1005 data disagree at the $90$ per cent level.}
\label{cold_abs_evolution}
\end{figure}

Fig.~\ref{cold_abs_evolution} shows the confidence contours in the power-law
slope $\Gamma$ and the column density \nhcold\ of an intrinsic neutral absorber,
as measured by the \rosat\ PSPC, \bepposax and the three \xmm\ observations.
Interestingly, the fixed-\nhcold\ but varying-$\Gamma$ scenario
(Table~\ref{xmm_cold}(d)), is marginally consistent with all the data-sets. The
spectral variations can be explained through variations in the photon index with
a fixed absorber column density of \hbox{$\sim1.4\times10^{22}\pcmsq$}. These
variations, however, do not seem to be correlated with changes in the
normalization of the power-law (which is presumably indicated by the
\hbox{$2$--$10\kev$} lightcurve in Fig.~\ref{lightcurves}(c)).

\begin{figure}
\rotatebox{270}{
\resizebox{!}{\columnwidth}
{\includegraphics{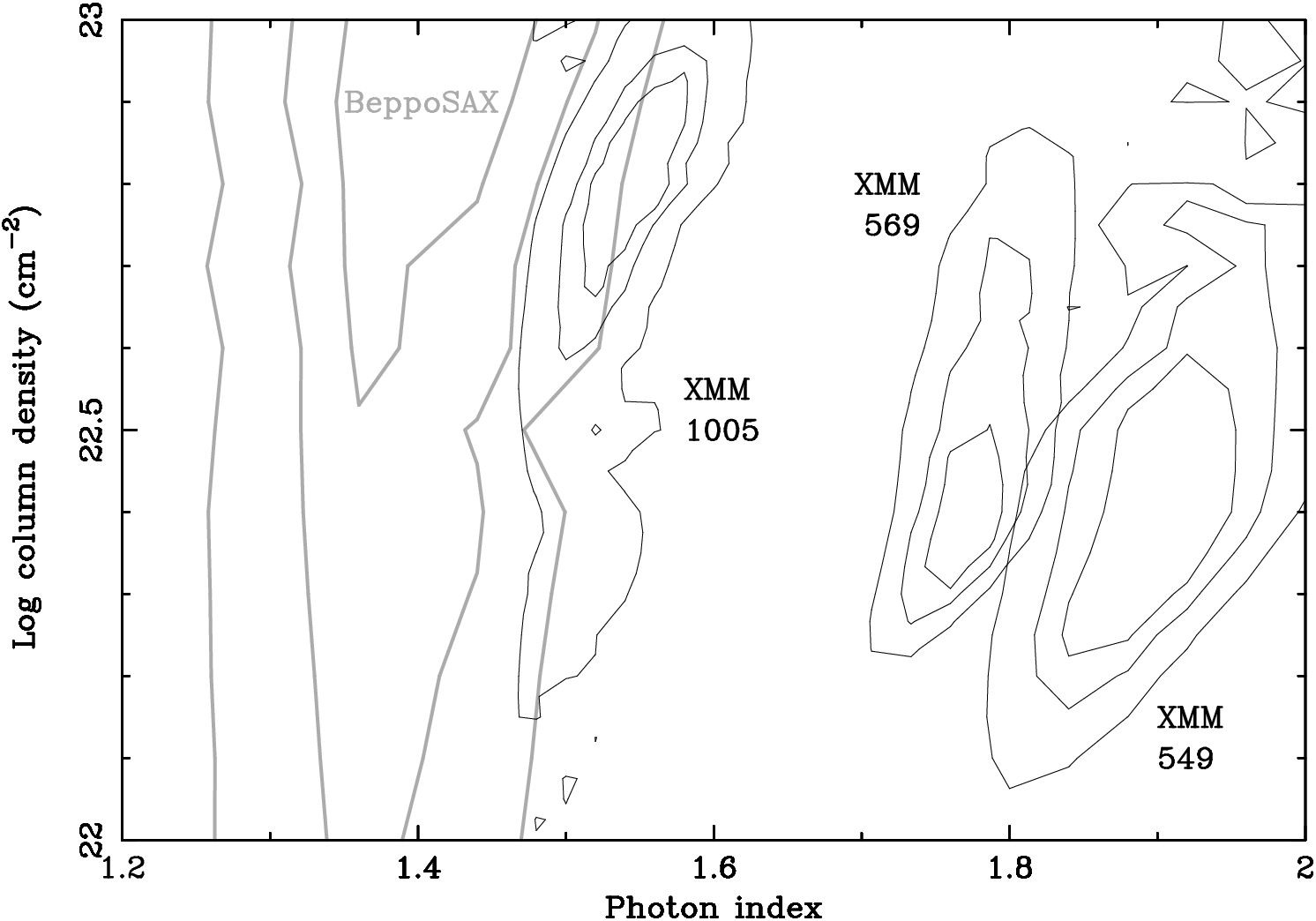}}}
\caption{Confidence contours ($68$, $90$ and $99$ per cent) in the photon index
  $\Gamma$ and the warm absorber column density \nhwarm. Contours are also shown
  for the \bepposax\ data \citep{fabian01b} and the data from \xmm\ revolutions
  549, 569 and 1005 (\citealp{worsley04b}; this paper). A fixed-\nhwarm\ warm
  absorber can account for all four observations through strong variations in
  the power-law slope and ionization parameter.}
\label{warm_abs_evolution}
\end{figure}

The spectral variation can also be explained successfully in the warm absorber
scenario. The most self-consistent situation is that of a fixed-\nhwarm\ but
varying-$\Gamma$ and varying-$\xi$, i.e. the column density of the absorber
remains fixed whilst the ionization state increases with the hardness of the
irradiating power-law; indeed, $\xi$ seems to increase with the hardening
$\Gamma$ from revolutions 549, 569 and 1005 (Table~\ref{xmm_warm}). This trend
can also be visualised in Fig.~\ref{warm_abs_evolution}, which shows the X-ray
spectrum hardening between the data from \xmm\ revolutions 549, 569 and 1005.  A
fixed \hbox{$\log_{10}(\nhwarm/\rm{cm^{-2}})\sim22.5$} absorber is consistent
with all the data-sets, with the underlying power-law softening from
\hbox{$\Gamma\sim1.4$} to $\sim1.9$, and then hardening back to $\sim1.5$ during
the \bepposax\ and the three \xmm\ observations. A trend is also seen in the
ionization parameter, which decreases from
\hbox{$\log_{10}(\xi/\rm{erg~cm~s^{-1}})\sim1.5$} to \hbox{$\sim0.8$}, and then
increases to \hbox{$1.6$}.

\section{Conclusions}

\begin{itemize}

\item We have observed an exceptional flare in the \hbox{$15.2\ghz$} radio
  lightcurve of \gb1428. The flux density has grown by a factor of
  \hbox{$\sim3$} in a rest-frame timescale of only \hbox{$\sim4$} months. The
  radio lightcurve also shows intra-day variations on the \hbox{$\sim10$} per
  cent level, consistent with previous observations.

\item Target-of-opportunity X-ray observations with \xmm\ do not see any
  evidence of a flare in the X-ray lightcurve, although the source continues to
  show \hbox{$\sim10$--$25$} per cent variability on rest-frame timescales of
  hours to days.

\item The ToO observations are able to confirm the unusual X-ray spectrum of the
  source, which is well-described by a hard (\hbox{$\Gamma\sim1.4$--$1.9$})
  power-law with a large amount (\hbox{$\sim10^{22}$--$10^{23}\pcmsq$}) of
  intrinsic absorption \citep[refer to][]{worsley04b}. The present data are
  unable to discriminate between a cold (neutral) or warm (ionized) absorption
  (a third possibility, of a spectral break in the power-law continuum, is
  discussed in detail in our previous paper).

\item Previous \xmm\ observations hinted at spectral variations in the source.
  These have been confirmed with the ToO data, which reveals that the photon
  index of the X-ray power-law has hardened from \hbox{$\Gamma\sim1.7$--$1.9$},
  to \hbox{$\Gamma\sim1.5$}. Interestingly, there is no clear evidence for a
  change in the column density of the absorption in either the cold or ionized
  absorber scenario. The spectral variations, including all previous data from
  other telescopes, are consistent with a scenario in which the column density
  of an ionized absorber remains fixed whilst the ionization state increases
  with the hardening of the power-law emission.

\end{itemize}

\section{Acknowledgments}
Based on observations with \xmm, an ESA science mission with instruments and
contributions directly funded by ESA Member States and the USA (NASA). MAW
acknowledges support from PPARC. ACF thanks the Royal Society for support. The
National Radio Astronomy Observatory is a facility of the National Science
Foundation operated under cooperative agreement by Associated Universities, Inc.
The authors would also like to thank the referee for helpful comments and
suggestions.

\bibliographystyle{mnras} 
\bibliography{mn-jour,MF1473rv}

\end{document}